\documentclass[aps,preprint,epsfig,showpacs,amsmath]{revtex4}
\usepackage{graphicx}
\begin{document}
\title{A simple model for the relative stabilities of DNA hairpin
structure} 
\author{Debaprasad Giri$^\dagger$, Sanjay Kumar and Yashwant Singh}
\affiliation{Department of Physics, Banaras Hindu University,
     Varanasi 221 005, India \\
$^\dagger$Centre for Theoretical Studies, IIT, Kharagpur
721 302, India}
\date{\today}
\begin{abstract}
A model of self-avoiding walk with suitable constraints on
self-attraction is developed to describe the conformational
behaviour of a single stranded short DNA molecule that form
hairpin structure. Using exact enumeration method we calculate
the properties associated with coil-hairpin transition.
Our results are in qualitative agreement with the experiment.
\end{abstract}
\maketitle

The structural and dynamic behaviour of a single stranded 
DNA molecule that form a stem-and-loop (hairpin) structures
in solutions is a subject of current interest \cite{1,2,3,4}. 
This is because DNA hairpin structure is known to participate in 
many biological functions, such as the regulation of gene 
expression \cite{5}, DNA recombination \cite{6} and facilitation
of mutogenic events \cite{7}. The DNA hairpin-containing 
domains may provide potential binding sites for exogenous 
drugs and endogenous proteins \cite{8}. It can also be used 
as DNA biosenser ({\it e.g.} molecular beacons) \cite{9}.

The single stranded DNA that form a stem-and-loop structures
under a given set of solution conditions are formed from 
synthetic oligonucleotides. For example, the oligonucleotide
$5'-${\bf CCCAA} - $(X)_m$ - {\bf TTGGG}$-3'$, where $X$ is any one
of the nucleotides and $m$ its number, is such a molecule 
which can easily be synthesized. Due to thermal fluctuations 
such a molecule may acquire different conformations. In a 
simplified description all of the configurations can be divided 
into two main states; the open and the closed one as shown 
in Fig. 1. The closed hairpin structure is stabilized due to 
pairing of complimentary bases at the two arms of the molecule
when hydrogen bonds are formed between them. The open state 
has high entropy due to the large number of configurations
achieved by a single stranded DNA chain. The closed-to-open 
transition requires sufficiently large energy to unzip all
the base pairs, whereas the closing transition requires the 
two arms of hairpin to come close to each other in space so 
that hydrogen bonding between the complimentary nucleotides
can take place. Experimentally one finds the rate of closing 
depends strongly on the properties of the hairpin loop, such 
as the length and rigidity, whereas the rate of opening is 
relatively unaffected by these properties \cite{1}. 

\begin{figure}
\includegraphics{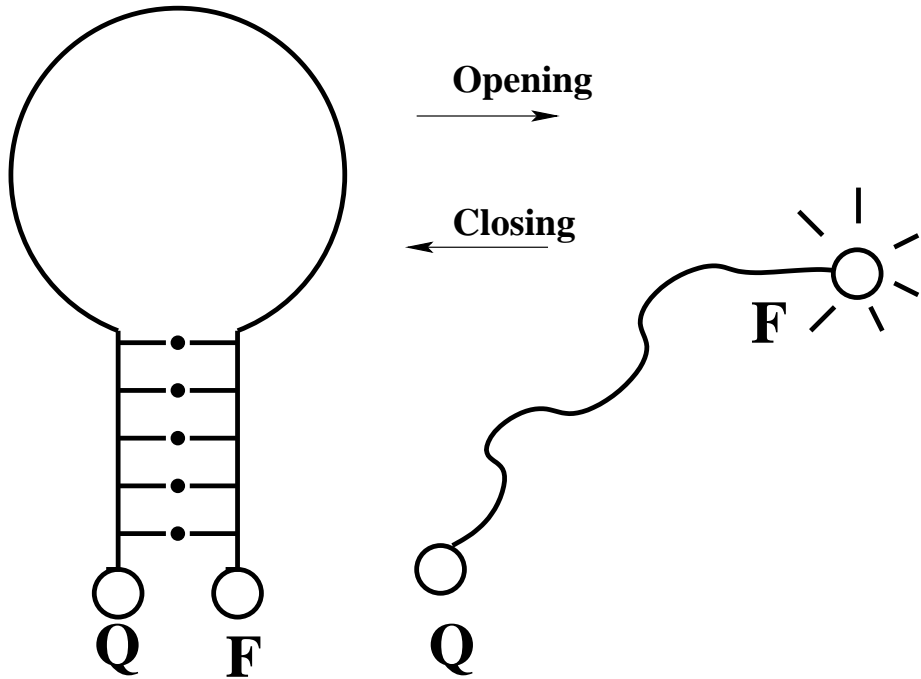}
\caption{Schematic of the conformational fluctuations of DNA 
hairpin-loop. This molecular beacon can fluctuate between
{\it open} and {\it closed} states in solution at ambient 
temperature.}
\end{figure}

In this note we describe a model and show that it explains 
some of the conformational behaviour associated with the 
formation of hairpin structure. 

We represent a linear polymer chain of $N$ monomers by a 
self-avoiding walk of $(N-1)$ steps on a two dimensional 
square lattice. Visited lattice sites represent the monomers
of the chain. The first $n$ sites of the walk represent 
nucleotides, say, {\bf A} or {\bf C} and the last $n$ sites 
the complimentary nucleotides, {\bf T} or {\bf G}. The remaining 
$(N - 2n)$ = $m$ sites represent the nucleotides of one type 
(either {\bf A} or {\bf T} or {\bf C} or {\bf G}). The repulsion 
at short distance between monomers ({\it i.e.} excluded volume)
interaction is taken into account by the condition of self
avoidance. When the complimentary monomers in a walk occupy 
nearest sites in a particular direction (say, clock wise), they
form a pair (known as base pair). The energy associated with 
this pair is $\epsilon$ ($\epsilon < 0$). The base paring cannot
take place if the nucleotides approach to each other in any other
way.  This condition has been imposed to take into account
the fact that in a single strand DNA chain the pairing
between complementary nucleotides takes place only when
they approach each other directly without the sugar
phosphate strand coming in between as shown in Fig. 2a.
Fig. 2b shows the situation in which the sugar phosphate strand 
is in between and therefore pairing can not take place.

\begin{figure}
\includegraphics{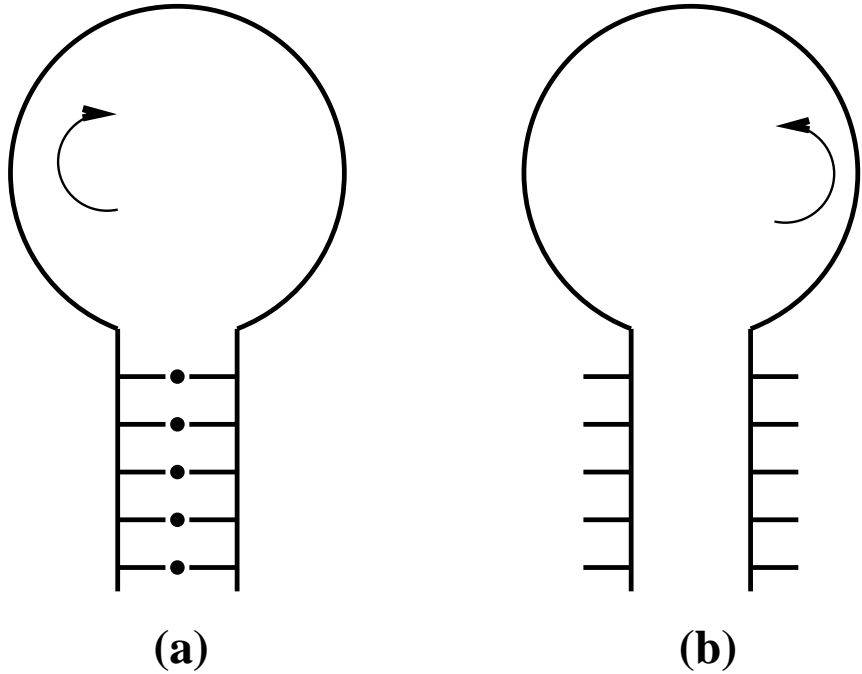}
\caption{Schematic of configurations showing the (a) formation and 
(b) non-formation of base pairs.}
\end{figure}

All possible conformations of walks of N sites mapped by a self-
avoiding walk are generated using exact enumeration technique 
\cite{10,11}.  Since the time involved in enumerating these 
conformations increases as $\mu^N$ where $\mu$ is the connectivity 
constant of the lattice, we have restricted ourselves to 
$12 \le N \le 22$.

The partition function of our interest is 
\begin{equation}
Z_N = \sum_{p=0}^{n} C_N(p) (e^{-\epsilon/k_B T})^p
\end{equation}

where $p$ is the number of base pairs and $C_N (p)$ 
the total number of configurations corresponding to 
walk of $(N-1)$ steps with $p$ number of pairs (base
pairs). The Helmholtz free energy of the system is given as
\begin{equation}
F = -k_B T \ln Z_N (T) 
\end{equation}
where $k_B$ is the Boltzmann constant and $T$ the temperature 
of the system. From Eq. (2) one can calculate the specific 
heat and entropy using the following relations:
\begin{subequations}
\begin{eqnarray}
C_v &=& - T \left ( \frac{\partial^2 F}{\partial T^2}\right )\\
S &=& - \left ( \frac{\partial F}{\partial T} \right )
\end{eqnarray}
\end{subequations}
In our calculation we have taken $n=5$ and $\epsilon = -0.08$ eV.
Dependence of the free energy ($F$) on temperature for $N= 22$ is 
shown in Fig. 3a. 

\begin{figure}
\includegraphics{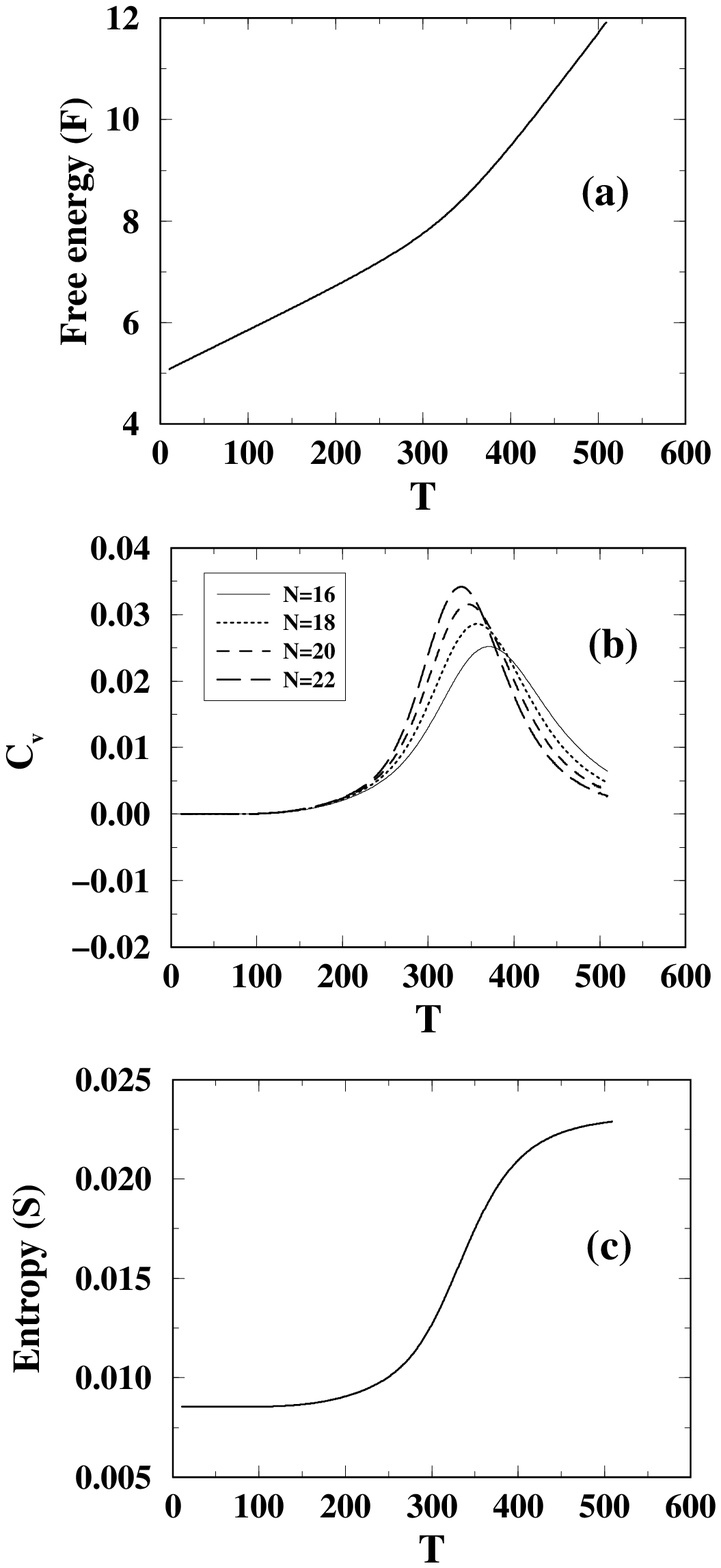}
\caption{(a) Plot of free energy ($F$) as a function of
temperature ($T$) for $N=22$; (b) Specific heat ($C_v$) 
as a function of temperature ($T$) for different $N$;
(c) Entropy ($S$) vs temperature ($T$) for $N=22$.}
\end{figure}

From this figure it is clear that there is a ``discontinuous" 
change in the free energy at $T = 338^o$K. In Fig. 3b we plot 
the specific heat as a function of temperature for several 
values of $N$. This curve has a peak suggesting a transition 
from the hairpin structure to the coil (or open) state. For 
$N = 22$, the peak is at $337\pm2^o$K. Since we are describing 
a finite system, the transition is not sharp; it takes place 
over an interval of temperature. This interval is obvious 
from the width of the peak in $C_v$ and also from  Fig. 3c
in which we plot entropy ($S$) as a function of $T$ for $N = 22$. 
The entropy is low at low temperature as the molecule is in a 
hairpin structural form whereas  at high temperature the entropy 
is higher as the molecule in coil form assesses a large 
number of conformations.

The partition function defined by Eq. (1) has six terms corresponding
to six values of $p$ from 0 to 5 ($p = 0$ corresponds to open state
of the chain whereas $1 \le p \le 4$ corresponds to partially closed 
state and $p = 5$ the final state of hairpin structure. We can 
find the probability of each of these states in the following way. 
\begin{equation}
P_p (T) = \frac{Z_p}{Z}
\end{equation}
where $Z = \sum_{p=0}^5 Z_p$

We compare $P_p(T)$ for different values of $p$ in Fig. 4a. 
It shows that the dominating structure are either all closed
or all open. 

The other quantity of interest is the fraction of open to
closed structures at a given temperature. The fraction of all 
open structure can be derived from the relation 
\begin{equation}
\chi_{-} = \frac{Z_0}{Z_0+Z_c}
\end{equation}
where $Z_0$ is for $p=0$ and $Z_c = \sum_{p=1}^5 Z_p$ and the 
fraction of closed structures
\begin{equation}
\chi_{+} = \frac{Z_c}{Z_0+Z_c}
\end{equation}
From Eq. (5) and (6) we get
\begin{equation}
\frac{\chi_{-}}{\chi_{+}} = \exp \left [ -\frac{1}{k_B T} 
(F_0 - F_c) \right ]
\end{equation}
where $F_0 = -k_B T \ln Z_0$ and $F_c = -k_B T \ln Z_c$

The fraction of all closed structure at a given temperature 
is found from the relation 
\begin{equation}
\chi_{+}^{\prime} = \frac{Z_5}{Z_0+Z_c}
\end{equation}

From this figure [Fig. 4b] one may note that while $\chi_{-}$ 
and $\chi'_{+}$ intersects at  $T = 337 \pm 2^o$K
in agreement with the coil-hairpin transition 
temperature,  $\chi_{-}$ and $\chi_{+}$ intersects at  
$T = 366 \pm 2^o$K. As shown in Fig. 4a the probability
of partially connected conformations are small and
confined to narrow temperature region. One may therefore
conclude that the conformational properties of this
kind of single DNA strand is primarily controlled by
either all open or all closed states.

\begin{figure}
\parindent -0.5in \includegraphics{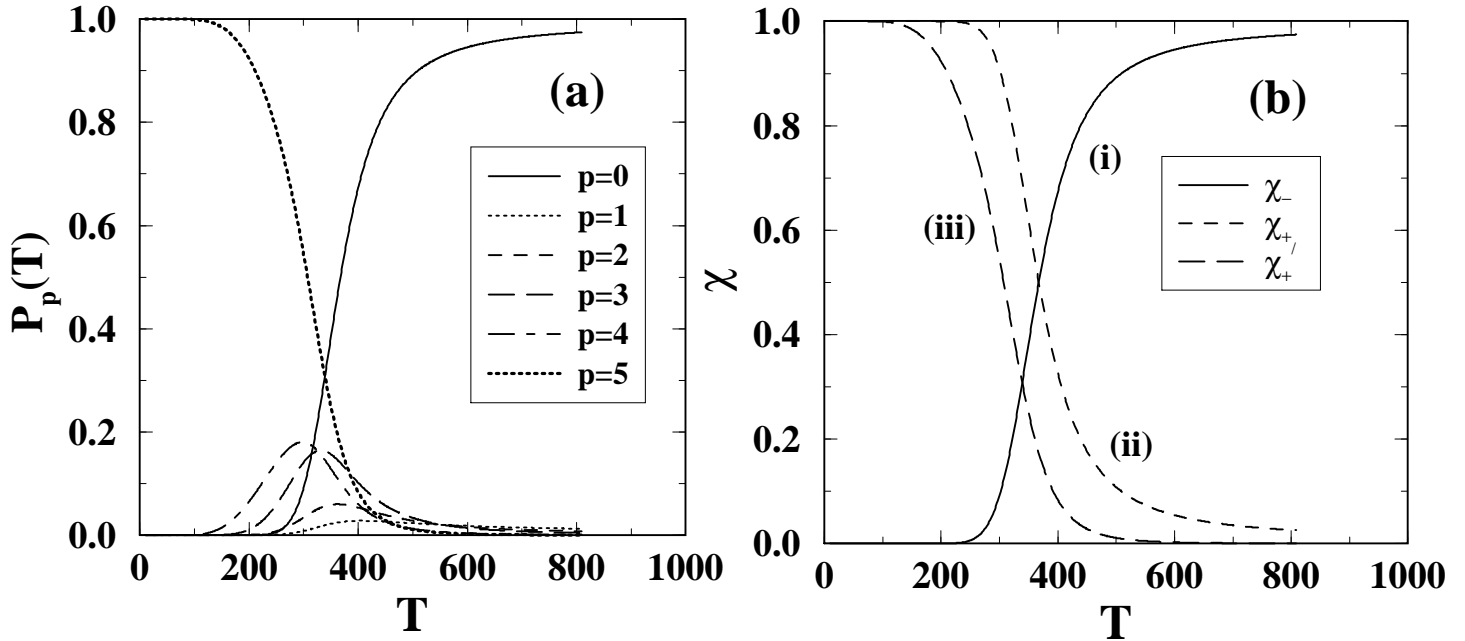}
\caption{(a) Probability of opening ($p=0$), closed ($p=5$) and
partially closed ($p=1,2,3,4$) states as a function of 
temperature ($T$); (b) (i) Fraction of all open structure 
($\chi_{-}$), (ii) fraction of closed structure ($\chi_{+}$) 
and (iii) fraction of all closed structure ($\chi_{+}^\prime$) 
as a function of temperature ($T$).}
\end{figure}

By varying the value of $\epsilon$, one can study the 
effect of base pair interaction on coil-hairpin transition. 
The change in $\epsilon$ value from -0.08 eV to -0.05 eV 
leads to the decrease in transition temperature by $100^o$K.

In conclusion, we have studied a simple model for DNA hairpin
loop structure and their relative stabilities.  We use lattice 
models of polymers along with exact enumeration technique to 
study the stability of  different conformations of DNA hairpin 
loop and the corresponding hairpin-coil transition.  Our results 
are in qualitatively agreement with the reported experimental 
work by Bonnet et al \cite{1}.  

\begin{acknowledgements}
We thank Navin Singh for many helpful discussions. 
Financial assistance from $INSA$, New Delhi and DST, New Delhi
are acknowledged.
\end{acknowledgements}

\end{document}